\documentstyle[aps,prc,preprint,tighten]{revtex}  
 
\begin{document}
\preprint{KSUCNR-07-96, nucl-th 9612020}
\title{Feedback from Freeze-out in Hydrodynamics\thanks{
 Dedicated to G.~Marx and K. Nagy on the occasion of their
		70$^{th}$ birthdays
}}
\author{
John J.~Neumann,\thanks{
E-mail: neumann@scorpio.kent.edu
}\
Boris Lavrenchuk,\thanks{
E-mail: boris@cnr2.kent.edu
}\ 
and George Fai\thanks{
E-mail: fai@ksuvxd.kent.edu  
} }

\address{
Center for Nuclear Research\\
Department of Physics\\
Kent State University\\
Kent, OH, 44242, USA\\
}
\maketitle 
\begin{abstract}
Most hydrodynamical calculations used in heavy-ion
physics ignore the effect of freeze-out matter carrying energy and
momentum away from the expanding fluid. In a simple one-dimensional model
we compare calculated energy density and velocity profiles, with and without 
interaction between fluid-like and freeze-out
parts of the system, in order to estimate the
importance of this effect.
\end{abstract}

\section{Introduction}
 
With new data from CERN's Superconducting Proton Synchrotron (SPS)
and with the expected completion of the Relativistic Heavy Ion Collider 
(RHIC) in Brookhaven in 1999, nuclear physicists will be able to 
look at higher temperatures in larger space-time volumes than
ever before. Of particular interest is the identification of
signatures of the transition to the high-energy phase of nuclear matter,
commonly referred to as the quark-gluon plasma (QGP).
It should be easier to measure the properties of such a transition 
with heavier ions because of the larger
times and volumes occupied by the resulting hot and dense
matter. With larger amounts of matter produced, we can hope that a 
description in terms of smoothly-varying locally equilibrated quantities
such as temperature and fluid-dynamical velocity is
reasonable, making hydrodynamical models especially valuable
in this scenario. The point of view 
where the physics is described in terms of such average properties 
rather than in terms of individual particle motions 
has obvious advantages.
The savings in computer simulation time allow research into a larger
parameter space and a wider variety of initial conditions can be explored.
Particle production calculations are routinely carried out in the
framework of such models. However, these models make a drastic
approximation when assuming a sudden change from fluid-dynamical
behavior to free-streaming (the usual effective definition of ``freeze-out'').
It should be kept in mind that, since our particle detectors register
free-streaming particles far from the interaction zone, a transition 
in the description from continuous matter variables to free particles is
unavoidable if the model is to have any predictive power or relevance
to measured data. The underlying question of better interfacing the
hydrodynamical and kinetic descriptions at freeze-out is studied by
several groups\cite{bernard,bug,grassi,scott}.

In traditional hydrodynamical models, all the matter in the system of interest
is assumed to be a thermalized fluid which
expands according to idealized hydrodynamics.
Within the fluid lies an isothermal surface which is
at the ``freeze-out temperature'', $T_{fo}$.
While the freeze-out surface plays no role in influencing the hydrodynamic
evolution, one assumes, for purposes of calculating particle spectra,
that matter outside the freeze-out surface is no longer in thermal equilibrium, 
i.e. particle interactions cease, and particles that cross the freeze-out
surface are free to move unimpeded to the detectors.
If, for example,  lepton pairs are measured, the decay of the freeze-out
particles can contribute a significant portion
of the observed leptons, and their spectra have interesting
properties which are sensitive to the hydrodynamic evolution\cite{jn2}.

The trajectory of the freeze-out surface is typically found by assuming that 
all the fluid is thermalized, even at temperatures below $T_{fo}$.
On the other hand, calculations of particle spectra
assume that particles near the freeze-out surface leave the fluid.
The energy and momentum carried away by these
freeze-out particles can be sizable, and
should properly be taken into account in a self-consistent calculation. 
The fact that some freeze-out particles may return to the fluid
through the moving freeze-out surface should also be accounted for. 
There has been some recent analytical investigation of how to do this by 
Bugaev\cite{bug}, who  derived equations which describe 
the trajectory of the freeze-out surface. Here we take the point 
of view of evolving a system from a given point in time
within the context of a computer simulation which includes the feedback
from this dynamical freeze out.
A Godunov-type calculation\cite{blol} appears well-suited for this endeavor. 
In this paper we show how to extend the Godunov method to 
handle the evolution of both thermalized fluid and 
freeze-out matter, and we compare to calculations of expanding matter
assuming all thermalized fluid (traditional hydrodynamics).
We demonstrate the effect on a one-dimensional example using the simplest
possible relativistic equation of state (EOS) for the thermalized fluid,
that of massless pions. These simplifying assumptions can be relaxed in
the future. Note that at the highest energies presently in the focus of 
attention, where a very large number of pions is created, 
even this most simplified study may be relevant to the interpretation of data.

The paper is organized as follows. In Section \ref{godmeth},
 we describe the essence of
Godunov-type methods as typically used today and prepare the way for the
inclusion of dynamical freeze-out. The details of how this is 
done are given in Section \ref{godfrz}, with the technicalities
 relegated to the Appendix.
Section \ref{genalg} summarizes the generalized algorithm implemented by our
computer code. Section \ref{results} contains the results,
 and we conclude in Section \ref{conc}.

\section{Godunov Method}\label{godmeth}

One successful approach to solving hydrodynamics on a computer is to use a
so-called ``Godunov-type'' method\cite{blol}.
In such a method, the space-time continuum is discretized into cells, i.e. the
algorithm calculates the changes in the average properties
of spatial cells as it makes finite steps in time.
Because conservation laws tell us how quantities (such as 
energy, momentum, and particle densities) in a cell change due to the flows
of those quantities through the cell ``walls'' during a timestep,
it only remains to determine what the flows should be to give the
correct hydrodynamical behavior. 

There are two common approaches within the framework of Godunov-type
methods to calculating the flows
between thermalized cells, an attempt to exactly calculate the intercell 
flow shape, and the approach of approximating the flows by 
uniform regions between cells.
Both methods assume that the cells can be treated as 
semi-infinite and uniform at the beginning of a given timestep.
This is reasonable insofar as the timestep is kept small enough
that the developing flow patterns do not extend past the middle
 of the cell, i.e. $\delta t < \delta x/2$, where $\delta t$
and $\delta x$ are, respectively, the timestep size and cell width$^1$.
Once this is assumed, it follows that neither a length nor
a time scale exists for the development of a flow pattern between two
cells. Therefore a similarity pattern, i.e. a pattern
that depends only on the dimensionless variable $\chi \equiv
x/t$, develops. This is also a reasonable approximation in more complex
geometries  as long as the similarity pattern
is much smaller than the size of the system as a whole.

In the most ambitious approach, one attempts to find the actual
shape of the flow pattern, e.g. the energy density $e(\chi)$ and the 
fluid velocity $v(\chi)$ across the cell boundary. These ``Riemann solutions''  
can be derived from the continuity equations for the relevant quantities
(in this case, energy and momentum) and the 
equation of state (EOS). In the case of a simple EOS, 
analytical solutions are possible\cite{baym,neumann}. 
For a realistic EOS, the shape of the flow pattern is found by numerical
integration. The alternative approach proposed by Harten, Lax, and 
van Leer, and completed by Einfeldt\cite{HLLE}, is 
commonly known as the ``HLLE'' method\cite{schneider}. Here
the flows between thermalized cells are approximated by uniform regions
of length proportional to the signal velocities (sound speed relativistically
added to fluid velocity).
HLLE saves considerable CPU time because the task of
calculating the actual flow patterns by numerical integration
is replaced by choosing approximate signal velocities,
which by conservation laws determine the conserved quantities
between the two receding discontinuities.
Discussion of the options for and choice of the signal velocities 
can be found in Ref.~\cite{schneider}. In the present work we follow
the HLLE algorithm as described there. 
We emphasize that in any Godunov-type
method the calculation of the flows between cells is at the heart 
of the fluid-dynamical problem.

For our purposes, a Godunov-type method is especially well suited,
because the Godunov equations (see eq. (\ref{1d}) below)
used to evolve a cell in time
are derived directly from conservation laws. Thus the validity of these 
equations is not restricted to ideal fluids, but 
they are true in general. Energy-momentum conservation is expressed in
 terms of the continuity equation for the energy-momentum tensor, 
\begin{equation}\label{contin}
\partial_{\mu} T^{\mu \nu} = 0 \,\, .
\end{equation}
The Godunov equations are obtained by averaging the time derivatives over
a cell, and for a system that varies in only one spatial
dimension, e.g. $x$, take the form
\begin{eqnarray}\label{1d}
\frac {d}{dt} \left[\int_{\delta x} dx \, T^{00}(x) \right] &=&
T_-^{0x} - T_+^{0x}  \,\,\,  ,\\[12pt]
\frac {d}{dt} \left[\int_{\delta x} dx \, T^{0x}(x) \right] &=&
T_-^{xx} - T_+^{xx}  \,\,\,  , \nonumber
\end{eqnarray}
where $\delta x$ refers to the spatial extent of the cell,
and the subscripts $-$ and $+$ refer respectively to the lower and upper
spatial bounds of the cell. The variable 
$t$ refers to the time as measured in the frame of the cells, which can be 
considered to be fixed in space (the ``cell frame"),
 with matter moving between them in the course of a simulation.
{}From inspection of eqs.~(\ref{contin}) and (\ref{1d})
it is clear that $T^{0x}$ represents
the flow of  $T^{00}$ and  $T^{xx}$ is the flow of  $T^{0x}$.
The energy-momentum tensor $T^{\mu \nu}$ is defined, in the rest
frame of thermalized fluid, to be
\begin{eqnarray}
T^{\mu\nu}=\left[
\begin{array}{cccc}
e & 0 & 0 & 0 \\
0 & p & 0 & 0 \\
0 & 0 & p & 0 \\
0 & 0 & 0 & p 
\end{array}
\right] \,\,\, ,
\end{eqnarray}
i.e. $T^{00}$ is the energy density $e$,
 $T^{0x}$ is the momentum density,
and $T^{xx}$ is the pressure $p$ (flow of momentum).
When the fluid is given a boost $v$ in the $x$-direction, we have
\begin{eqnarray}\label{Teq}
E &\equiv& T^{00} = (e+p)\gamma^2-p\,\,\, ,\\
M &\equiv& T^{0x} = (e+p)v\gamma^2 \nonumber \,\,\, , \\ 
P &\equiv& T^{xx} = (e+p)v^2\gamma^2+p \,\,\, , \nonumber
\end{eqnarray}
where $\gamma^2 = 1/(1-v^2)$,
and we have given mnemonic names (reflecting the physical quantities to
which these tensor elements reduce in the rest frame)
$E$ (energy density), $M$ (momentum
density), and $P$ (pressure or flow of momentum density $M$)
 to the three tensor
elements of interest. ($E$ and $M$ are used in Ref. \cite{schneider}, 
$P$ is introduced in this work.) For concise treatment, we introduce
the notation $U_i: E,M,P$ for $i=1,2,3$. We also use 
$q_i: e,p$ ($i=1,2$) for the energy density and pressure in the
rest frame of thermalized fluid. The notation $F(U_i)$ for
the flow of quantity $U_i$ will also be used occasionally.

Recall that in a non-curved  geometry 
the flows are similarity patterns, and thus the values of 
$M(\chi=0)$ and $P(\chi=0)$ remain constant throughout
the timestep. It then follows that
\begin{eqnarray}
\Delta \left< E\right> &=& \frac {\delta t}{\delta x}
\left(M_- - M_+\right),\\[12pt]
\Delta \left< M \right> &=& \frac {\delta t}{\delta x}
\left(P_- - P_+\right),
\end{eqnarray}
where $\left< T^{\mu \nu}\right>$ signifies the average value of
 $T^{\mu \nu}$ in the cell.
This is the form that the computer code actually uses to 
evolve a cell from one timestep to the next.
In the examples presented in this work we use a 
simple ultra-relativistic
equation of state, so particle number conservation is ignored
and energy-momentum conservation as presented above is sufficient.

\section{Generalized Godunov Method Including Freeze-Out}\label{godfrz}

We wish to extend the Godunov method to handle freeze-out
as well as thermalized fluid. Strictly speaking, we are not
extending the Godunov method {\it per se}, but simply using it in a way
that exploits its generality. The general problem as we have mapped it
out consists of solving for flows in three cases: 
between two cells of thermalized fluid, 
between two cells of non-thermalized (freeze-out) matter, and 
between one cell of each kind.
For this last case, the freeze-out surface 
lies somewhere in one of the two cells, and for simplicity, we 
approximate the flow with the flow due to thermalized matter 
(approximation with non-thermalized matter is an equally valid choice). 
This can be done because we 
expect the density gradient to be well-behaved in the region
connecting thermalized fluid with freeze-out matter, as justified 
{\it a posteriori} by our results (Sect.~5). 

The physics of freeze-out can be described in terms of matter 
that was originally thermalized and treated as a fluid 
expanding to such a low density that large mean-free
paths make fluid behavior an invalid approximation. In 
fluid-dynamical models of nuclear collisions, such as ours, this is assumed 
to happen at a well-defined ``freeze-out temperature'' below which the
particles which comprise the matter lose thermal contact.
In relativistic nuclear collisions $T_{fo} \approx 100-150$ MeV,
and the freeze-out particles are expected to reach the
detectors without any further interaction.
In freeze-out matter it no longer
makes sense to calculate flow patterns based on fluid behavior, or 
even to assume that local quantities such as energy density ($e$) and
pressure ($p$) are related by an equation of state. 
However, since the Godunov equations do
nothing more than express the conservation of energy and 
momentum in a discretized system, they can still be applied,
as long as the correct values for the flows, 
$F(E)=M$ and $F(M)=P$, (and in our case, $F(P)$, the flow of 
$P$) are used. It should be pointed out that there is no 
relativistically covariant continuity
equation for $P$, meaning that pressure is
not conserved like energy  and 
momentum. However, in weakly interacting matter, such as
freeze-out, one can still calculate the contribution 
to the pressure due to a particle of given momentum, so one can calculate
the ``flow of $P$'' by considering the rate at which particles cross
the cell wall. Thus, in the case of freeze-out matter, we assume 
$\Delta \left<P\right> = (\delta t/\delta x)
\left[F(P)_- - F(P)_+\right]$ (the calculation of $F(P)$ is described
in the Appendix), and for thermalized fluid we calculate $P$ from
$E$, $M$, and the EOS.

In order to calculate flows of freeze-out matter, it is necessary to
keep track of more than $E$ and $M$. This is because for
non-thermalized matter, which does not follow an equation of state, 
$P$ does not 
have an implicit relationship to $E$ and $M$. In principle,
one should keep track of a number of degrees of freedom of the
order of the number of particles in the matter. However, including
 just tensor element $P$ (in addition to the usual $E$ and $M$) 
in the description captures the essence of freeze-out
behavior, which is that the free-streaming 
particles tend to move at a single velocity,
making $P_0$ (the value of $P$ in the zero-momentum frame, where $M$=0,
referred to as the ``matter rest frame'') 
smaller than would be expected for thermalized
matter with the same $E_0=e$. Thus our calculation can 
be thought of as a first order
approximation of the coupling effect. 

If the potential energy is negligible compared to the 
kinetic and rest mass energies,  e.g. in a non-interacting pion gas,
it is natural to express the elements 
of the energy-momentum tensor in terms of the distributions of particles
\cite{deGroot}:
\begin{equation}\label{kin}
T^{\mu \nu}= \int \frac{d^3 p}{p^0} p^{\mu}p^{\nu} f\left(
\frac{p^0}{T} \right),
\end{equation}
where $f\left(p^0/T \right)$ describes the distribution
of the particles.
For thermalized particles viewed in the matter rest frame, as
defined above, the function
$f\left(p^0/T \right)$  is the Bose-Einstein or 
Fermi-Dirac distribution (Bose-Einstein for the pion gas considered here).
The momentum distribution of particles immediately inside the freeze-out
surface is known, because we have made the assumption of 
thermal equilibrium everywhere inside the freeze-out surface. 
Because we assume a sharp freeze-out surface, the particles crossing
the surface will retain the momenta 
they had while still in the thermalized fluid. We will use (\ref{kin}) 
for the calculation of the elements of the energy-momentum tensor in
non-thermalized matter. The only remaining freedom is in specifying the
distribution function, 
which is described in detail in the Appendix. 

We assume massless pions for the freeze-out particles.
This is because pions are the lightest, and therefore most abundantly
produced hadrons, so our model is closely linked to actual 
experimental data at the highest beam energies available.
Massive particles could be used at a price of some
complication of the EOS for thermalized fluid and of the calculation of
velocity distributions. Here we limit ourselves to an extremely simple
case to illustrate the method, which we expect to find
more general applications (e.g. the nonrelativistic scenario of evaporation 
from a fluid). 

\section{Structure of the Algorithm}\label{genalg}

Here we summarize the tasks performed by the generalized 
Godunov algorithm for the above simplified scenario at each timestep. 
Before evolving the system in time, the cells are first initialized with
values for $v$, the  cell frame matter velocities,  and $q_i$, 
the rest frame thermodynamic quantities ($e$ and $p$). 
After all the cells are initialized,
the cell frame tensor elements $U_i$ ($E$, $M$ and $P$) are calculated.
The piecewise-constant distribution is replaced by a piecewise-linear
distribution, the slopes being calculated from the values of $U_i$
in neighboring cells. Thus  two distinct 
values of $U_i$ that correspond to two adjacent cells are associated 
with each boundary (see \cite{schneider} for details). The difference
between these two $U_i$ determines a flow at a given boundary. These
flows give rise to second order accuracy in time 
by allowing the calculation of half-timestep quantities, including the
signal velocities and flows used in the HLLE scheme.
The new second-order flows determine 
the evolution of $U_i$ inside each cell (throughout the code, if the
$U_i$ are changed, the corresponding $q_i$ and $v$ are updated immediately
afterward). Finally, $e$ and $v$ for each cell are output.
At this point one time step is completed. 

For cells with freeze-out matter,
we depart from the standard HLLE scheme, as described in \cite{schneider}.
If a cell is determined to be below the freeze-out temperature
(based on the rest-frame energy density $e$), the flow is calculated 
assuming free-streaming particles.
To do this, we first need to parametrize the velocity distribution in terms
of the tensor elements $U_i$, as described in the Appendix.
To make a calculation with the freeze-out coupling turned off, one
simply uses the standard HLLE to calculate all the flows regardless of 
energy density.

\section{Results}\label{results}

We ran most of our simulations with $\delta t$=0.03 fm for the timestep
and $\delta x$=0.1 fm for the cell size. The dependence on the timestep 
was tested, and we found that decreasing $\delta t$ to 0.01 fm did not 
alter the results appreciably. We use the equation of
state of a thermalized massless boson gas (pions). The initial temperature 
was $T_0=200$ MeV for -2.7 fm $\leq x \leq$ 2.7 fm, and vacuum 
everywhere else. Initial velocities were all set to 
zero. Because of the symmetry about $x=0$, only one side of the system
($x \geq 0$) is shown in the figures.
The freeze-out temperature was chosen to be $T_{fo}=140$ MeV, close to the 
mass of a pion. (This appears to be a realistic estimate of the 
freeze-out temperature\cite{jn2}.) The corresponding isotherm is called the 
freeze-out isotherm and defines the freeze-out surface. 
In a calculation of particle production, particles that cross the freeze-out
surface are assumed to lose thermal contact and move freely 
to the detectors. Therefore, the trajectory of the freeze-out surface
is important, and is our primary concern in this work.
The results shown were obtained by a second order calculation in time,
as described in Section~4.
For our simple problem, with a ``well-behaved'' EOS,
we get nearly identical results for first and second
order simulations; for equations of state with a phase transition, or
other systems where macroscopic discontinuities may develop, 
a second order treatment is necessary to get accurate results.

\begin{figure}[t]
\vspace*{7.0cm}
\includegraphics{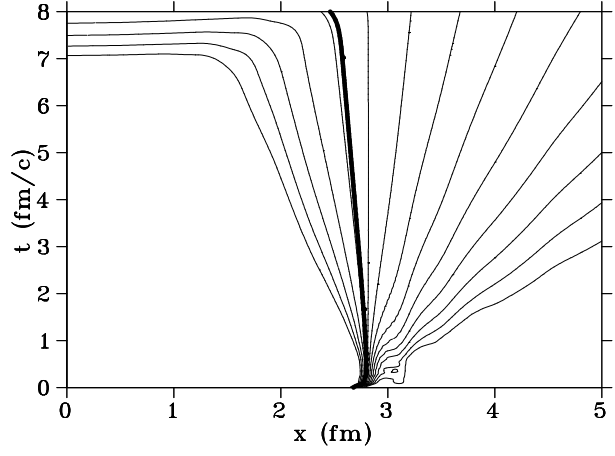}
\caption[]{Lines of constant energy density, for the case without 
coupling between the freeze-out and thermalized fluid.
The thick line ($e$=49 MeV/fm$^3$) is the freeze-out isotherm.}
\label{fig1}
\end{figure}

\begin{figure}[b]
\vspace*{7.0cm}
\includegraphics{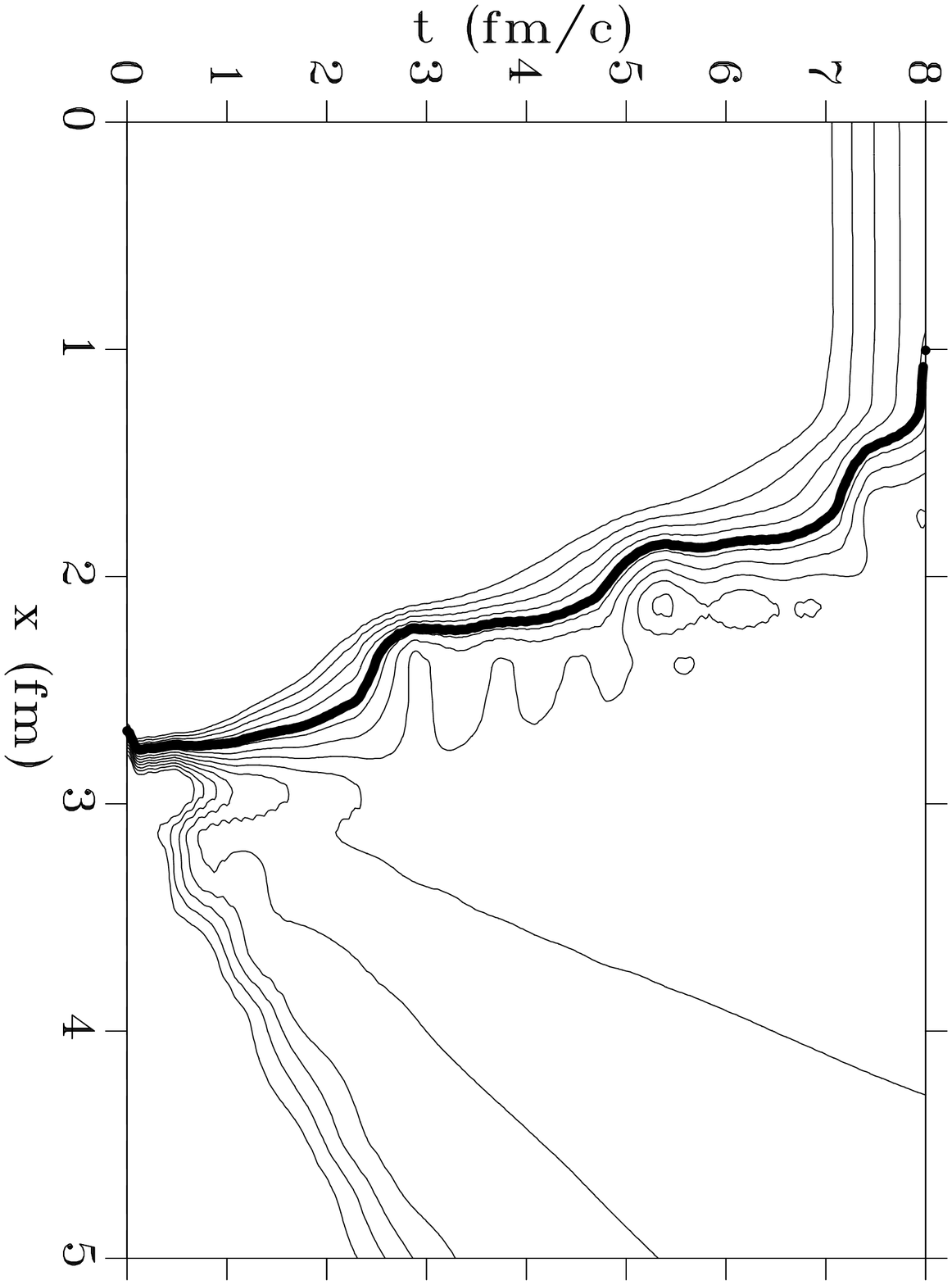}
\caption[]{Lines of constant energy density, for the case with 
coupling  between the freeze-out and thermalized fluid.
The thick line ($e$=49 MeV/fm$^3$) is the freeze-out isotherm.}
\label{fig2}
\end{figure}

Figures \ref{fig1} and \ref{fig2} are contour plots of
 $e$ in $x$-$t$ space for the
evolution of a finite one-dimensional slab expanding into
the vacuum in both directions; only one side, $x \geq 0$, is shown,
without and with feedback from freeze-out, respectively.
The isotherms, starting from the right side of the diagram, represent
energy densities from 5 to 70~MeV/fm$^{3}$, in increments of 
5~MeV/fm$^{3}$. The thick line is the $e$=49~MeV/fm$^3$ isotherm,
which corresponds to the freeze-out temperature ($T_{fo}=140$~MeV), 
or freeze-out surface. 
We see that it has a distinctly different
slope in $x$-$t$ space when we include the coupling between thermalized fluid
and freeze-out. The increased inward velocity of the freeze-out surface
would result in more freeze-out particles being measured, because the number
is sensitive to the relative velocity between the matter and the freeze-out
surface. 

The contour plot for $e$ also demonstrates that there are large regions of
small density gradient in the freeze-out matter
in the case of freeze-out coupling. Nearly the
entire region outside the freeze-out surface is at 
around half the freeze-out energy
density. This is in contrast to the case without freeze-out coupling, which 
displays a much smoother change of the energy-density profile. 
In the case with feedback from freeze-out, there are also much larger
fluctuations in energy density, notably a region of lower
density (a ``bubble'') that seems to have formed 
in the freeze-out matter around $t=5.0$ fm, 
but disappears at around $t=7.0$~fm. 
Whether this is indicative of new physics or
merely a result of the uneven flow caused by 
the crudeness of our choice of velocity distributions, or the
limited number (three) of moments of the momentum distribution we use,
needs further investigation.

\begin{figure}[t]
\vspace*{7.0cm}
\includegraphics{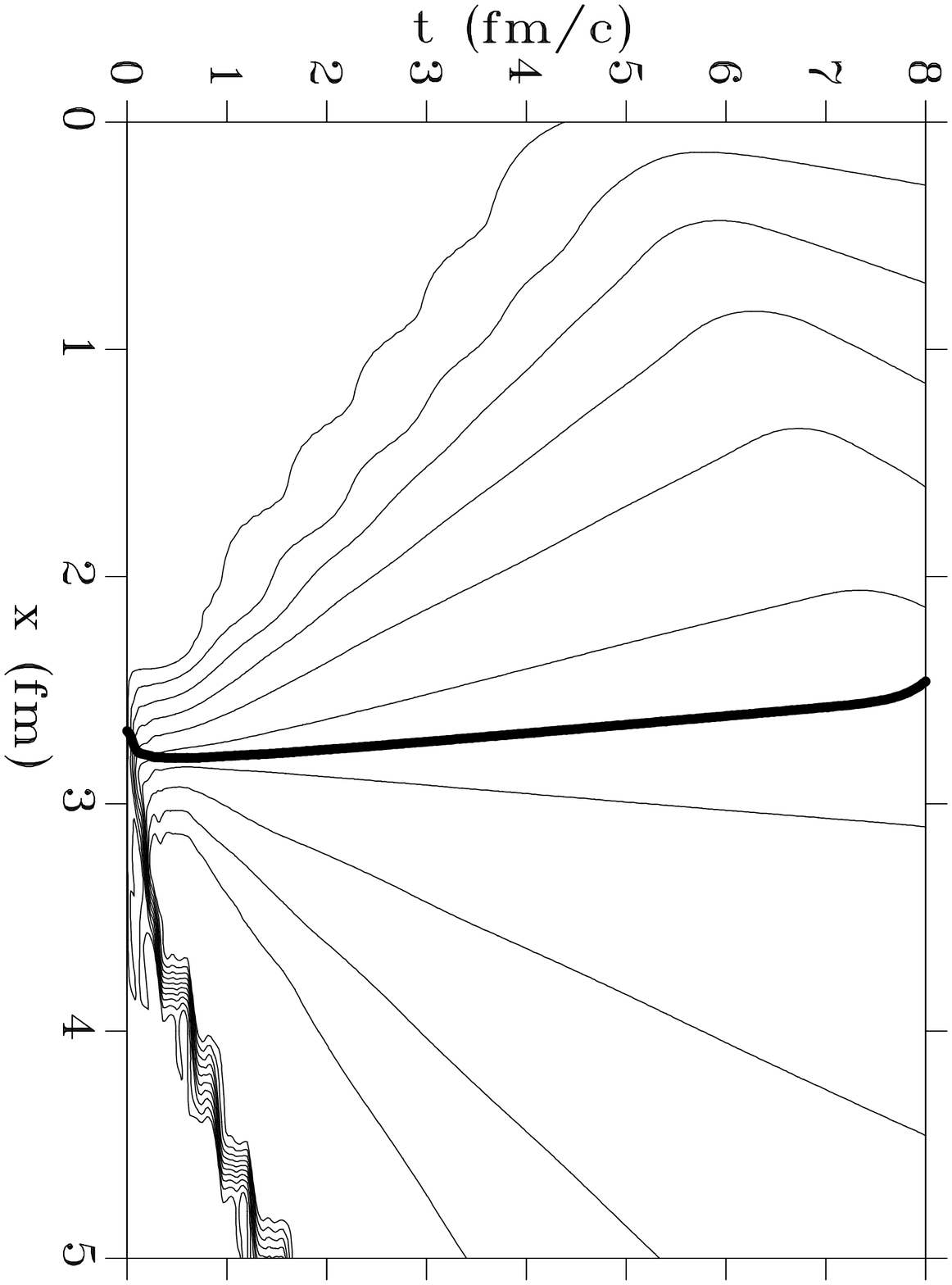}
\caption[]{Lines of constant velocity $v$, for the case without 
coupling  between the freeze-out and thermalized fluid.
The thick line is the freeze-out surface.}
\label{fig3}
\end{figure}

\begin{figure}[b]
\vspace*{7.0cm}
\includegraphics{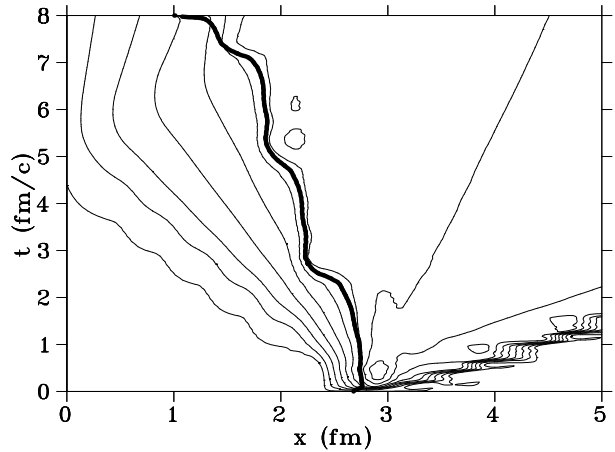}
\caption[]{Lines of constant velocity $v$, for the case with 
coupling  between the freeze-out and thermalized fluid.
The thick line is the freeze-out surface.}
\label{fig4}
\end{figure}

Figures \ref{fig3} and \ref{fig4}, 
the contour plots for $v$ in $x$-$t$ space,
without and with feedback from freeze-out, respectively,
corroborate our expectations and the results displayed in the $e$-plots.
In the case of freeze-out coupling (Figure \ref{fig4}),
 there are large regions outside 
the freeze-out surface that have little variation in velocity, 
indicating that the particles ``free-stream''
with no resistance. The contours are, from $x=0$ outward,
$v= 0.01, 0.1, 0.2, \cdots 0.8, 0.9$. 
The group of lines in the lower right corner is an artifact of the
contour plotting due to the fact that fast rarefied matter abuts
the vacuum, whose velocity is taken to be zero.
Most of the freeze-out matter moves outward at a speed around  0.65$c$
in the case when the feedback from the freeze-out is taken into account. 
The inhomogeneities that appear seem to be due to
disturbances that originate at or near the freeze-out surface, where
particles with a range of velocities cross into the freeze-out region.

\section{Conclusions}\label{conc}

We have generalized the conventional Godunov algorithm to consistently 
include freeze-out matter in fluid-dynamical simulations. In particular,
the energy and momentum carried away by the freeze-out particles and
the feedback from those particles is accounted for in our generalized
treatment. In the present paper we illustrated this dynamical freeze-out 
in a simple one-dimensional example. The net result of the dynamical 
treatment is a rapid eating into the thermalized fluid
as shown by the trajectory of the
freeze-out surface, as compared to the treatment without dynamical freeze-out.
Also, there is a more sudden change across the freeze-out surface
to freeze-out matter that is moving more or less uniformly, as expected
from free-streaming particles. In addition, the treatment with freeze-out 
coupling displays more fluctuations, partly due to our simplifying 
assumptions.

For a more realistic simulation we intend to use the generalized Godunov
algorithm in more complicated geometries and with more complicated 
equations of state. We expect the method to find
many applications. One of these may be to combine 
the improved freeze-out modeling with three-fluid
hydrodynamics\cite{3fluid}. Such a combination would be especially
useful for calculating production of particles that are sensitive to 
freeze-out (e.g. dileptons). One suspects that the effect of the
freeze-out/fluid coupling on the trajectory 
of the freeze-out hypersurface will influence pion rapidity
distributions at least as much as differences between
isothermal and isochronous freeze-out\cite{bernard}.
Photon spectra are also expected to be modified, albeit to a lesser extent.

We believe that fluid-dynamical calculations with an improved treatment of
freeze-out will be very competitive with transport-theoretical
modeling. This is partly because of great savings in CPU time and partly
due to the physics insight afforded by a treatment dealing with average
quantities, as opposed to running computer simulations
with a very large number of chance events.  
 
\section{Appendix}

In our approach each cell is characterized by the three tensor elements
$E$, $M$ and $P$. These are Lorentz-transformed into the frame where
$M$, the momentum density, is zero. This is referred to as 
the zero-momentum frame,
or ``matter rest frame'', and we will denote the tensor elements in this frame
by $E_0$, $M_0=0$ and $P_0$. The velocity of this frame relative to the frame
of the cells is the ``matter velocity'' $v$. 
In the matter rest frame, the cells are characterized by $E_0$ and $P_0$.
For weakly interacting matter in general, 
$E_0=e$, the energy density, and $P_0=p$,
the pressure. For the simple (ultra-relativistic) equation of state 
of our thermalized fluid, $P_0=E_0/3$. For freeze-out matter there is 
no relationship between $E_0$ and $P_0$ except
for constraints imposed by causality.

For thermalized massless particles, the momentum distribution
is isotropic, and $p^x\equiv p^0 \xi$, where we use the notation 
$-1 \le \xi \le 1$ for the dimensionless velocity
 (in terms of the speed of light) along the 
$x$-axis, and $p^0$ is the time-like component of the 4-momentum.
Therefore we can factor each integral in (\ref{kin})
into an integral over total energy and an integral over velocity
in the $x$-direction, e.g.
\begin{eqnarray}\label{E0int}
E_0 &=& \int \frac{d^3 p}{p^0} p^{0}p^{0} f\left(
\frac{p^0}{T} \right)\\
&=& \int_{-1}^{+1} d\xi  \, 2\pi \int_0^{\infty} dp^0 \, 
\left( p^0 \right)^3 \,
f\left( \frac{p^0}{T} \right) \nonumber \\
&\equiv& \int_{-1}^{+1} d\xi \, h(\xi),\nonumber
\end{eqnarray}
where $h(\xi)$, the energy distribution with respect to velocity,
is obviously a constant $E_0/2$ due to the fact that the integral
over $p^0$ is independent of $\xi$ for thermalized massless particles.
(Note that here the subscript on $E_0$ refers to the matter frame,
while the superscripts on $p$ refer to components
of the 4-momentum.) Due to 
this factorization for massless particles we can
recast the momentum distribution as
a velocity distribution along the $x$-axis.
For other equations of state, $h(\xi)$ would be defined,
but not necessarily a constant.
In a similar manner, we obtain:
\begin{eqnarray}\label{otherint}
M_0 &=& \int_{-1}^{+1} d\xi \, \xi \, h(\xi)=0,\\
P_0 &=& \int_{-1}^{+1} d\xi \,\xi^ 2 \, h(\xi) \,\,\, .\nonumber
\end{eqnarray}

For non-thermalized matter, we parametrize $h(\xi)$ subject to the conditions 
that it be (1)  even with respect to $\xi=0$,
which guarantees $M_0=0$, and (2) non-negative. 
For cells near thermal equilibrium, i.e. $P_0 \approx E_0/3$, we
want the simplest function that differs gently from a constant and
still satisfies the above criteria; the choice
$h(\xi)=a\xi^2+b$ ($a$ and $b$ constant) works well,
where $a=(45 P_0 -15 E_0)/8$ and $b=(9 E_0 -15 P_0)/8$.

For much of the freeze-out matter,
$P_0$ is too small (namely, $P_0 < E_S \equiv E_0/5$) 
to satisfy $h(\xi)=a\xi^2+b \ge 0$ for all $-1 \le \xi \le +1$. 
In this case we interpolate between
a function of the form $a\xi^2+b$ and a delta function, $(E_0/2)\delta(\xi)$.
The use of the delta function reflects our assumption that the particles
in freeze-out matter tend to move along together at nearly the same
velocity.
Occasionally it is necessary, for large $P_0$ ($> E_L \equiv 3E_0/5$), 
to use a double delta function 
of the form $h(\xi)=(E_0/2)[\delta(-d)+\delta(d)]$, where
$d$ is a velocity. In this case the matter is described better 
by a superposition of movement at a couple velocities than 
by homogeneous particle movement.

To calculate the flows of matter over the right-hand (left-hand)
cell boundary, we calculate
the integrals (\ref{E0int}) and (\ref{otherint})
for particles with cell-frame velocities $v(\xi)$ greater than
(less than) zero, where $v(\xi)=(\xi+v)/(1+\xi v)$.
For example, the flow of energy $F(E)$ (due only to matter in the cell being
considered) over the right-hand side of
the cell is 
\begin{eqnarray}
F(E)=M^{\prime}_+ &=& \int_{-v}^{+1} d\xi \, \frac{dM}{d\xi}\\
&=& \int_{-v}^{+1} d\xi \,\gamma^2 \frac{d}{d\xi}\left( 
v E_0+(v^2+1) M_0 + v P_0 \right)\,\,\, \nonumber
\end{eqnarray}
where $M^{\prime}_+$ is the momentum density, viewed in the cell frame, 
 due to particles with velocity
between 0 and 1. The derivatives of the rest frame quantities are simply
\begin{eqnarray}
\frac{dE_0}{d\xi}&=& h(\xi),\\
\frac{dM_0}{d\xi}&=& \xi\, h(\xi),\nonumber \\
\frac{dP_0}{d\xi}&=& \xi^2\, h(\xi) \, \, \, . \nonumber
\end{eqnarray}
The evaluation of the flow of $P$ requires the numerical integration of
\begin{eqnarray}
F(P) &=& \int_{-v}^{+1} d\xi \,v(\xi) \, \frac{dP}{d\xi}\\
&=& \int_{-v}^{+1} d\xi \,v(\xi) \, \gamma^2 \frac{d}{d\xi}\left( 
v^2 E_0+2v M_0 +  P_0 \right)\nonumber
\end{eqnarray}
where 
$\gamma^2=1/(1-v^2)$ is the gamma factor due to the relative velocity
$v$ between the rest frame and the cell frame.

Flows from two neighboring cells are added together to determine the
total flow across the shared interface. 
These flows are then used by the code to update $U_i$ at the 
next full timestep.
 
\vskip 20pt
\noindent \large {\bf Acknowledgement}\normalsize
\vskip 10pt
\noindent
The authors thank Dirk Rischke for valuable discussions, 
which helped them recognize the importance of this work, 
and for pointing out K.~Bugaev's paper. Useful conversations with Scott 
Pratt are also acknowledged. This research was supported in part by the 
U.S. Department of Energy under Grant No. DOE/DE-FG02-86ER-40251.
\section*{Notes} 
\begin{itemize}
\item[1.] We use the convention $\hbar=c=1$.
\end{itemize}

\end{document}